\documentclass[aps,prd,superscriptaddress,groupedaddress,nofootinbib,longbibliography,preprint]{revtex4-1}
\newcommand{\bew}{\begin{widetext}}
\newcommand{\enw}{\end{widetext}}
\newcommand{\bee}{\begin{equation}}
\newcommand{\ene}{\end{equation}}
\newcommand{\bea}{\begin{eqnarray}}
\newcommand{\ena}{\end{eqnarray}}
\newcommand{\beal}{\begin{align}}
\newcommand{\enal}{\end{align}}
\newcommand{\beald}{\begin{aligned}}
\newcommand{\enald}{\end{aligned}}
\newcommand{\bes}{\begin{subequations}}
\newcommand{\ens}{\end{subequations}}

\def\to{\rightarrow}

\def\d{{\rm d}}

\def\call{\mathcal{L}}
\def\calm{\mathcal{M}}

\def\calo{\mathcal{O}}

\def\mev{\,{\rm MeV}}
\def\gev{\,{\rm GeV}}
\def\tev{\,{\rm TeV}}

\def\fb{\,{\rm fb}}

\def\ab{\,{\rm ab}}

\usepackage{amsmath,amssymb,amsfonts}
\usepackage[colorlinks]{hyperref}
\usepackage{graphicx}
\usepackage{subfigure}
\graphicspath{{./pic/}}

\def\call{\mathcal{L}}
\def\calm{\mathcal{M}}

\def\calo{\mathcal{O}}

\def\to{\rightarrow}

\def\d{{\rm d}}
\def\tev{{\rm TeV}}
\def\gev{{\rm GeV}}

\begin{document}


\title{Constraining Gluonic Contact Interaction of a Neutrino-philic Dark Fermion at Hadron Colliders and Direct Detection Experiments}

\author{Kai Ma}
\email[]{kai@xauat.edu.cn}
\affiliation{Faculty of Science, Xi'an University of Architecture and Technology, Xi'an, 710055, China}
\affiliation{Department of Physics, Shaanxi University of Technology, Hanzhong 723000, Shaanxi, China}
\author{Lin-Yun He}
\email[]{a1164432527@gmail.com}
\affiliation{Faculty of Science, Xi'an University of Architecture and Technology, Xi'an, 710055, China}
\affiliation{Department of Physics, Shaanxi University of Technology, Hanzhong 723000, Shaanxi, China}

\date{\today}

\begin{abstract}
Weakly interacting fermions with the Standard Model particles are promising candidates for the dark matter. In this paper, we study signatures of the gluonic interactions of a dark fermion and a neutrino at hadron colliders and direct detection experiments. The lowest order interactions are described by contact operators in dimension 7. At hadron colliders, the mono-jet production is the most sensitive channel. And these operators can also induce both spin-independent and spin-dependent absorption of the dark fermion at nuclear targets. We show that for a nearly massless dark fermion, the energy scales are constrained to be higher than 500 GeV and 1.2 TeV by the current LHC and HE-LHC searches, respectively. Furthermore, we also find that almost all the parameter space accessible by the spin-independent absorption has been excluded by the current LHC constraints. In contrast, for spin-dependent absorption at light nuclear targets there is still some parameter space which cannot be reached by current and upcoming LHC searches.
\end{abstract}

\maketitle

\tableofcontents

\setcounter{page}{1}
\renewcommand{\thefootnote}{\arabic{footnote}}
\setcounter{footnote}{0}

\section{Introduction}
\label{sec:intro}

The Dark Matter (DM), a non-luminous and massive matter component of our universe, has been shown by both astrophysical and cosmological measurements to possess a significant fraction of the whole contents.
\cite{Bertone:2004pz,Young:2016ala,Arbey:2021gdg,Fairbairn:2022gar}. 
Unfortunately, so far, besides the fact that the DM is massive, due to the absence of direct observations of the DM, we know very little about its other physical properties.
We can only confirm that it should be stable and neutral so that its relic abundance is consistent with the current experimental observations \cite{Arguelles:2023nlh,Tuominen:2021wrl,Chadha-Day:2021szb,deSalas:2020hbh}. One of the most well-motivated DM candidates is the Weakly Interacting Massive Particle (WIMP) \cite{Roszkowski:2017nbc,Pospelov:2007mp,Arcadi:2017kky,Giagu:2019fmp,Schumann:2019eaa}, which has been extensively studied both theoretically and experimentally \cite{Cui:2015eba,Bernabei:2000qn,Steigman:1984ac,Bernabei:1996vj}. However, to date, there has been no positive experimental result from any of the three approaches: collider searches \cite{Argyropoulos:2021sav,Lorenz:2019pxf,Boveia:2018yeb,Penning:2017tmb,Kahlhoefer:2017dnp}, direct detections \cite{Billard:2021uyg,DelNobile:2021wmp,Misiaszek:2023sxe,Cebrian:2021mvb,Liu:2017drf,MarrodanUndagoitia:2015veg}, and indirect observations \cite{Leane:2020liq,Slatyer:2021qgc,deLaurentis:2022oqa,deDiosZornoza:2021rgw,Das:2021hnk,Hutten:2022hud}. Therefore, it is a reasonable conjecture that DM interacts with Standard Model (SM) particles in an exotic manner.

Recently, absorption processes of the DM on nuclear and electron targets have received lots of attention because of the DM's promising signature in direct detection experiments, particularly for neutrino-philic interactions
\cite{Dror:2019onn,Dror:2019dib,Dror:2020czw,Ge:2022ius,Li:2022kca,Ge:2023wye,Ma:2024aoc,Ge:2024euk,PandaX:2022osq,PandaX:2022ood,CDEX:2024bum}. 
The PandaX Collaboration reported that the limit on DM - nucleon interaction cross section for a fermionic DM with mass $40\,\mathrm{MeV}$ is $1.5 \times 10^{-50}\,\mathrm{cm}^2$ with 90\% $\mathrm{C.L.}$ \cite{PandaX:2022osq}. The PandaX \cite{PandaX:2022ood} and CDEX \cite{CDEX:2024bum} Collaborations also investigated the sub-MeV fermionic DM absorption at electron targets. For a DM with mass $5\,\mathrm{keV}$, the upper bounds of the absorption cross sections obtained by the CDEX experiment are $5.5 \times 10^{-46}\,\mathrm{cm}^2$ for vector and $1.8 \times 10^{-46}\,\mathrm{cm}^2$ for axial-vector mediation.
The interaction operators responsible for the DM-nuclei scattering can also be probed at hadron colliders \cite{Ma:2024aoc,Bishara:2017pfq,Belyaev:2018pqr,Dreiner:2013vla,Cepedello:2023yao,Roy:2024ear,Buchmueller:2014yoa,DEramo:2014nmf}. It was shown that for spin-independent (SI) absorption, the results from the LHC with center-of-mass energy $\sqrt{s} = 13\,\mathrm{TeV}$ and a total luminosity $139\,\mathrm{fb}^{-1}$ have excluded almost all the parameters accessible by the direct detection experiments \cite{Ma:2024tkt}.
In case of spin-dependent (SD) scattering of a DM with mass in the range $\sim [1, 100]\,\mathrm{MeV}$, absorption at lighter nuclear target (for instance the Borexino experiment \cite{Borexino:2018pev}) can give much stronger constraints than the LHC \cite{Ma:2024tkt}. It is clear that the collider searches and direct detections are complementary \cite{Gninenko:2023pkv,Boveia:2022adi,Alanne:2022eem,Chakraborti:2021mbr}.
However, the previous works mainly focused on four-fermion contact interactions involving a quark pair. Here we study the interaction operators involving a pair of gluons, a neutrino and a dark fermion $\chi$. We will study signatures of the gluonic operators at hadron colliders and direct detection experiments.

The rest of this paper is organized as follows. 
In Sec.~\ref{sec:eff}, we summarize our parameterization of the gluonic interaction 
operators and discuss invisibility of the dark fermion at high energy colliders.
In Sec.~\ref{sec:monoj}, we study constraints on the gluonic interaction operators
by mono-jet events at the current and future upgrades of the LHC. 
In Sec.~\ref{sec:AbsNT}, we explore absorption signals of the gluonic operators
at nuclear targets, and the spin-independent and spin-dependent absorption processes
are studied in subsections \ref{sec:DD:SI} and \ref{sec:DD:SD}, respectively.
Our conclusions are given in Sec.~\ref{sec:conclusion}.

\section{Effective Operators}
\label{sec:eff}
The four-fermion contact couplings of a dark fermion to quarks have been extensively studied. At hadron colliders, they can be searched for via mono-$X$ (the $X$ can be a photon, a $Z/W$ boson, or a jet) \cite{Abdallah:2015uba,daSilveira:2023hmt,Gabrielli:2014oya,Gershtein:2008bf,
Hicyilmaz:2023tnr,Kawamura:2023drb,Yang:2017iqh,Abdallah:2019tpo,No:2015xqa,Bell:2012rg,Alves:2015dya,Wan:2018eaz,Bell:2015rdw,Claude:2022rho,Belyaev:2018ext,Bai:2015nfa,Belwal:2017nkw,Bernreuther:2018nat,Krovi:2018fdr,Liew:2016oon,Bhattacharya:2022qck}.
For parton-level center-of-mass energy $\sqrt{\hat{s}}$ 
and the new physics scale $\varLambda$,
while the cross sections of four-fermion contact interactions 
grow as $\hat{s}/\varLambda^{4}$~\cite{Dror:2019onn,Dror:2019dib,Dror:2020czw,Ge:2022ius,Li:2022kca,Ge:2023wye,Ma:2024aoc,Ge:2024euk}, those of the gluonic operators grow as $\hat{s}^2/\varLambda^{6}$. The gluonic contribution has an additional suppression factor of $\hat{s}/\varLambda^{2}$. Nevertheless, at high energy colliders,
since $\hat{s}\sim\varLambda^{2}$ (or the upper bound on $\varLambda$ is around $\sqrt{\hat{s}}$), the reduction effect on the corresponding exclusion limits of the gluonic operators turns out to be negligible ($\hat{s}/\varLambda^{2}\sim1$).
In contrast, in direct detection experiments, because the center-of-mass energy is much smaller than the new physics scale, leading to a significant suppression of the cross sections, the suppression effect is very significant.
As a result, the combined exclusion limits can be very different for the gluonic operators and
the four-fermion operators.
The collider bounds on gluonic operators are comparable to those on four fermion operators, while the constraints from direct detection experiments are much weaker.
Here, we investigate the gluonic contact interactions of the dark fermion $\chi$ \cite{Bishara:2017pfq,Morgante:2018tiq}. 

In contrast to the quark-dark fermion couplings, at hadron colliders, the dominant dark-fermion production induced by the gluonic interaction is the mono-jet process \cite{Godbole:2015gma}.
Furthermore, it is usually assumed that the dark fermion couples to gluons in pairs \cite{Bishara:2017pfq,Morgante:2018tiq} as a result of some discrete symmetry such as $\mathbb{Z}_2$, which guarantees that the dark fermion is stable enough to be a  DM candidate. In this case, the signals can be probed by the elastic scattering of the dark fermion off a nuclear target in direct - detection experiments \cite{Hisano:2010ct}.
However, as long as the DM is light enough, its decay width can be small enough such that it can survive until today as a DM candidate \cite{Dror:2019onn,Dror:2019dib,Dror:2020czw,Ge:2022ius,Li:2022kca,Ge:2023wye}. 
In this paper, we focus on the gluonic contact interactions of a dark fermion and a neutrino. Similar to the four-fermion contact coupling involving a dark fermion and a neutrino \cite{Dror:2019onn,Dror:2019dib,Dror:2020czw,Ge:2022ius,Ma:2024aoc,Ge:2024euk,Li:2022kca,Ge:2023wye,PandaX:2022osq,PandaX:2022ood,CDEX:2024bum}, such operators can induce the absorption of the dark fermion at a nuclear target.
In the effective field theory (EFT) framework, the lowest-order interactions are given as the following two dimension 7 operators:
\begin{eqnarray}
\label{eq:ggo:cpe}
\mathcal{O}_1 
&=& 
\frac{\alpha_s}{12 \pi \varLambda_1^3}
(\bar{\chi} \nu) G^{a \mu \nu} G_{\mu \nu}^a
\\[3mm]
\label{eq:ggo:cpo}
\mathcal{O}_2 
&=& 
\frac{\alpha_s}{8 \pi \varLambda_2^3 }
(\bar{\chi} i\gamma_5 \nu) G^{a \mu \nu} \widetilde{G}_{\mu \nu}^a
\end{eqnarray}
where $\nu$ is the SM neutrino and $G^{a \mu \nu}$ is the gluon field strength; 
$\alpha_s(m_Z) = 0.12$ is the strong coupling constant used for normalization, and $\varLambda_{1,2}$ are the corresponding energy scales characterizing the strengths of these two operators.

The above EFT description is valid as long as the energy scale $\varLambda_i$ \cite{Dreiner:2013vla} (assuming $\varLambda_i$ represents a single energy scale here) or the mass of the possible mediator \cite{Busoni:2013lha,Busoni:2014sya,Busoni:2014haa} is higher than the net momentum transfer of the related transition.
Since our operators are defined at the TeV scale, when studying the signals of these EFT operators at a low-energy scale, such as in direct detection experiments, the running effect of the operators can be significant \cite{Hill:2011be,Frandsen:2012db,Vecchi:2013iza,Crivellin:2014qxa,DEramo:2016gos}.
It was shown that the running effect can result in a mixture of operators with different Lorentz structures at low energy \cite{Bishara:2017pfq}. However, such a mixing effect is only numerically relevant when the coupling between the dark fermion and the top-quark is significant \cite{Bishara:2017pfq}. 
In this paper, we assume that the coupling of the dark fermion to the top-quark is turned off, and hence the running effect is negligible. The interaction between the dark fermion and the top-quark can be searched for independently, and we will study this aspect in a separate work.

For the dark fermion $\chi$ to be a  DM candidate, a UV completion that realizes the above effective operators is necessary. Generally, the DM relic abundance can always be accommodated through a variety of thermal 
(and non-thermal) production mechanisms. As a result, the energy scales $\varLambda_{1,2}$ studied in this paper can be treated as completely free parameters. 
For example, our gluonic interactions can appear naturally in an axion (axion-like) portal model \cite{Fitzpatrick:2023xks,Dror:2023fyd,Anilkumar:2024tda} with a nontrivial mixing between the dark fermion $\chi$ and the neutrino \cite{Dror:2019onn,Dror:2019dib,Dror:2020czw,Ge:2022ius,Li:2022kca,Ge:2023wye,Ma:2024aoc,Ge:2024euk}.
Furthermore, a more comprehensive theoretical model inherently entails a larger set of parameters. This situation significantly impedes the feasibility of conducting a model-independent investigation.
Therefore, hereafter, we simply assume that the dark fermion $\chi$ only couples to the gluons of the SM particles and is described by the operators $\calo_{1,2}$.

\begin{figure}[h] 
\centering
\includegraphics[width=0.48\textwidth]{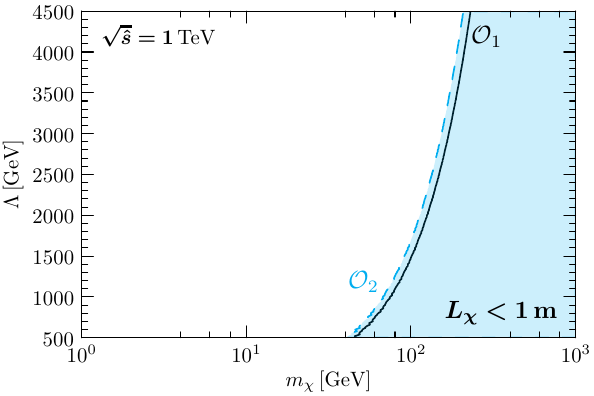}
\includegraphics[width=0.48\textwidth]{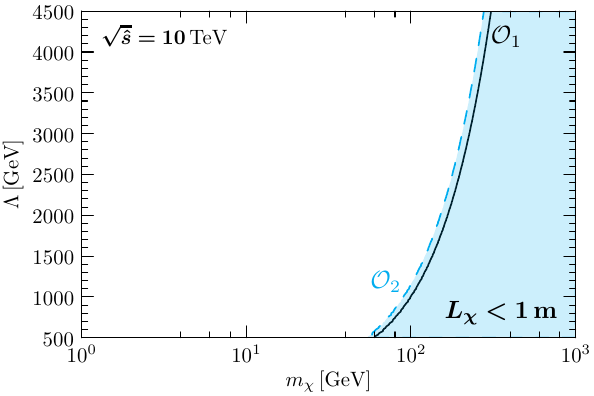}
\caption{\it 
Typical decay length of the dark fermion in the $m_\chi$-$\varLambda$ plane
for parton level center-of-mass energy $\sqrt{\hat{s}}=1\tev$ (\textbf{left}) and
$\sqrt{\hat{s}}=10\tev$ (\textbf{right}).
}
\label{fig:dcylen}
\end{figure}

On the other hand, an examination of the stability of the dark fermion is necessary, since we will treat the dark fermion as a completely invisible particle when we study the signal properties at the hadron collider. Within our assumption, the dark fermion can decay only through the following channel,
\bee
\chi \to \nu + g + g \,.
\ene
The total decay widths are given as,
\bea
\varGamma^{\chi \to \nu gg}_{G} 
&=&
\frac{ \alpha_s^2  m_\chi^7 }{ 34560 \pi^5 \varLambda_1^6 } \,,
\\[3mm]
\varGamma^{\chi \to \nu gg}_{\widetilde{G}} 
&=&
\frac{ \alpha_s^2 m_\chi^7 }{ 15360 \pi^5 \varLambda_2^6 } \,.
\ena
For a heavy dark fermion $\chi$, invisibility at the collider requires that it does not decay inside the collider. In the following analysis, we simply require that the typical decay length should be larger than 1 m. One can have an intuitive picture of the invisibility by referring to this point. At a collider with (parton level) center-of-mass energy $\sqrt{\hat{s}}$, the typical decay length of the dark fermion is given by:
\bee
L_{\chi} 
= \gamma_{\chi} \tau_{\chi} \beta_\chi
= \frac{ \beta_\chi \sqrt{\hat{s}} }{2 m_\chi \varGamma_{\chi} }  
\Big( 1 + \frac{ m_\chi^2 }{ \hat{s} } \Big)\,,
\ene
This formula describes the typical decay length of the dark fermion. Here, $\beta_\chi$ and $\gamma_\chi$ are the velocity and the relativistic boost factor of the dark fermion, respectively.
For a relatively light dark fermion ($m_\chi \sim 0$), its velocity can be approximated as $\beta_\chi \approx 1$. Thus, one has $L_{\chi}  \approx  \sqrt{\hat{s}}/(2 m_\chi \varGamma_{\chi} ) $.  
Fig.~\ref{fig:dcylen} shows the regions with $L_\chi < 1\,$m for typical center-of-mass energies at the LHC, $\sqrt{\hat{s}}=1\,\mathrm{TeV}$ at the left panel and
$\sqrt{\hat{s}}=10\tev$ at the right panel.

One can see that for a dark fermion with mass $m_\chi \lesssim 20\,\mathrm{GeV}$, the assumption of invisibility is always valid. For a relatively heavier dark fermion, the restriction of invisibility depends on the energy scale $\varLambda_i$.
We have to point out that this constraint is model dependent, rather than a general restriction, particularly when the dark fermion $\chi$ is just one component of a whole dark sector \cite{Gori:2022vri,Marra:2019lyc,Deliyergiyev:2015oxa,Hofmann:2020wvr,Lagouri:2022ier}.
If the dark fermion $\chi$ is light, say $m_\chi < \varLambda_{\rm QCD}$, the above decay channels are forbidden due to colour confinement. Hence, for a dark fermion with mass $m_\chi \lesssim 20\,\mathrm{GeV}$, one can safely treat it as an invisible particle at the high energy colliders.
On the other hand, for a heavier dark fermion, a whole model of the dark sector is necessary to interpret the dark fermion as a  DM candidate. Here we omit the details of such UV-completed models, and we focus on the model independent constraints on the signal operators at the hadron collider and direct detection experiments. One can easily convert our results to bounds on the UV-completed model parameters.

In addition, in consideration of the unique advantage of the high energy collider in detecting heavy dark particles, the full mass window of the dark fermion $\chi$ will be scanned in our studies related to collider searches. Only in Sec.~\ref{sec:AbsNT}, where the dark fermion $\chi$ is interpreted as a possible DM candidate, the mass of $\chi$ is restricted to be in the range $\sim (1, 100)$ MeV. 
Within this mass range, due to QCD color confinement, the direct decay channel is prohibited. In this regard, the dark fermion can be considered completely stable.
On the other hand, decay channels can be opened at the loop level. For example, the four-fermion contact interactions studied in Refs. \cite{Dror:2019onn,Dror:2019dib,Dror:2020czw,Ge:2022ius} fall into this category. However, the exact decay width of the dark fermion strongly depends on how the neutrino-philic property is realized at the ultraviolet scale. One such example is the Axion (Axion-like) portal model \cite{Fitzpatrick:2023xks,Dror:2023fyd,Anilkumar:2024tda}, which involves a non-trivial mixing between the dark fermion and the neutrino \cite{Dror:2019onn,Dror:2019dib,Dror:2020czw,Ge:2022ius,Li:2022kca,Ge:2023wye,Ma:2024aoc,Ge:2024euk}.
Given this situation, we did not aim to explore the decay details further for $m_\chi \lesssim 100$ MeV. Investigating how such effective operators can be reasonably incorporated into an ultraviolet completed model is beyond the scope of this work. We plan to address this aspect in future research.

\section{Mono-Jet Production at the LHC}
\label{sec:monoj}

At the hadron collider, the dark fermion is always produced 
in association with a SM neutrino. Hence, the kinematics can be very different from the pair production of $\chi$ (in case the gluons couple to the dark fermion in pairs \cite{Bishara:2017pfq,Morgante:2018tiq}), particularly when the dark fermion is heavy.
Assuming that the dark fermion is always invisible, the dominant production channel induced by the effective operators defined in \eqref{eq:ggo:cpe} and \eqref{eq:ggo:cpo} is the mono-jet process. Hereafter, our numerical results regarding the mono-jet production at the LHC are obtained using the toolboxes \textsf{MadGraph} \cite{Alwall:2014hca,Frederix:2018nkq} and \textsf{FeynRules} \cite{Alloul:2013bka}.
\begin{figure}[h]
\centering
\includegraphics[height=0.11\textheight]{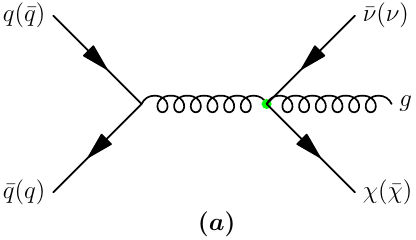}
\quad
\includegraphics[height=0.11\textheight]{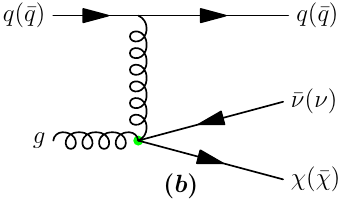}
\\[2mm]
\includegraphics[height=0.11\textheight]{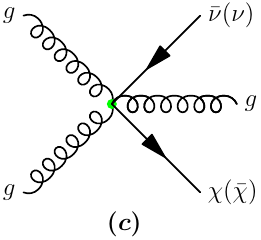}
\quad
\includegraphics[height=0.11\textheight]{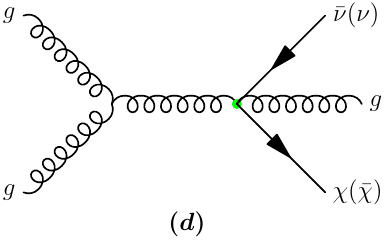}
\quad
\includegraphics[height=0.11\textheight]{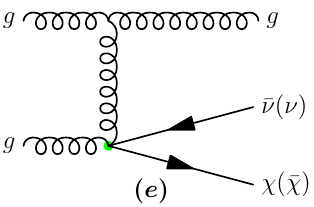}
\\[2mm]
\includegraphics[height=0.11\textheight]{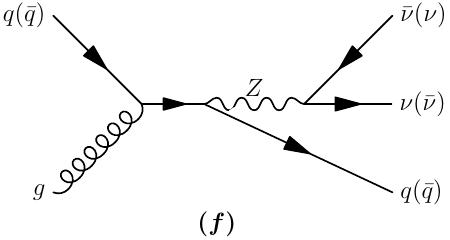}
\quad
\includegraphics[height=0.11\textheight]{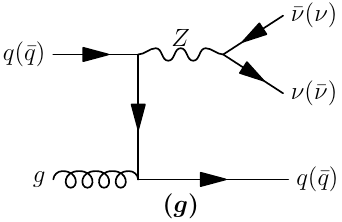}
\quad
\includegraphics[height=0.11\textheight]{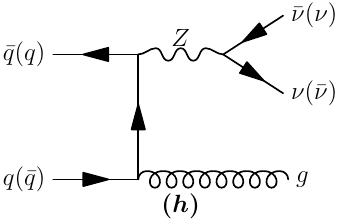}
\caption{\it Feynman diagrams contributing to the mono-jet events.
The plots \textbf{(a)}, \textbf{(b)}, \textbf{(c)}, \textbf{(d)} and \textbf{(e)} are for the signal 
process $pp \to \chi \bar\nu j$ and $pp \to \nu\bar\chi j$. The plots
\textbf{(f)}, \textbf{(g)} and \textbf{(h)} are for the irreducible background process 
$pp\to Z^\ast j \to \nu\bar\nu j$.
}
\label{fig:feyn:monoj:LHC}
\end{figure}
Fig.~\ref{fig:feyn:monoj:LHC} (a)-(e) show Feynman diagrams of the signal. One can see that both quarks and gluons can be the incoming partons. So relatively stronger constraints on the signal operators are expected. Fig.~\ref{fig:feyn:monoj:LHC} (f)-(h) represent Feynman diagrams of the irreducible background: associated production of a jet and a $Z$ boson with its subsequent invisible decays.

\begin{figure}[th]
\centering
\includegraphics[width=0.32\textwidth]{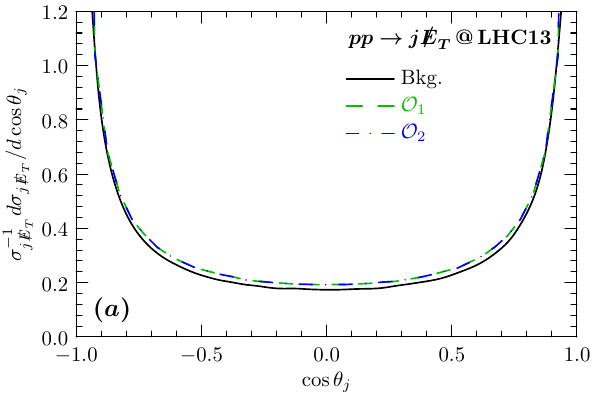}
\hfill
\includegraphics[width=0.32\textwidth]{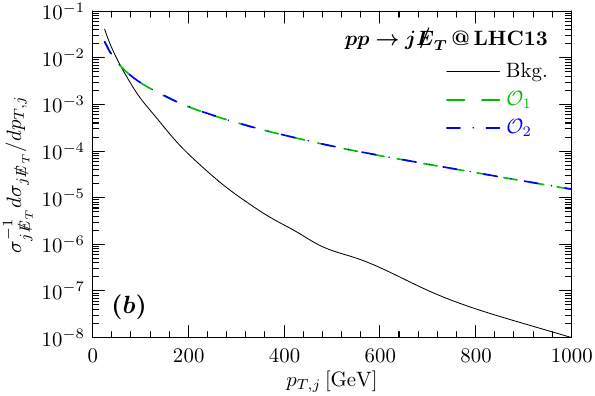}
\hfill
\includegraphics[width=0.32\textwidth]{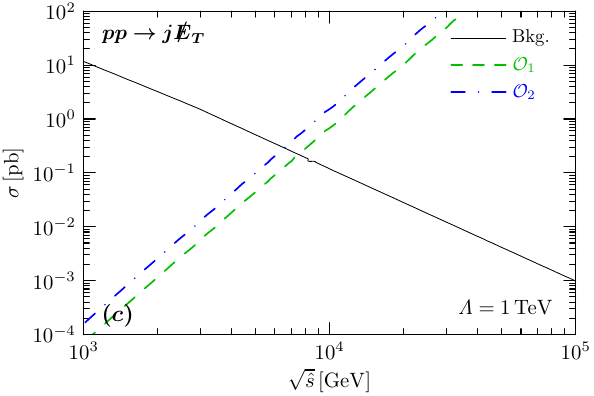}
\caption{\it 
Normalized parton level distributions of the polar angle ($\theta_j$) \textbf{(a)}
and transverse momentum ($p_{T,j}$) \textbf{(b)} of the jet
in the lab frame with center of mass energy $\sqrt{s} = 13\tev$. 
\textbf{(c):} total cross sections of the background and signals as functions of the
center-of-mass energy at parton level, $\sqrt{\hat{s}}$. 
In all the three panels, the signal (colorful non-solid curves) are shown for  a massless dark fermion and energy scale $\varLambda_i = 1\tev$, 
and the background (black-solid curve) stands for the irreducible
contribution from the channel $pp\to Z^\ast j \to \nu\bar\nu j$.
}
\label{fig:ca:pt:mj:LHC13}
\end{figure}

For both the signal and irreducible background, the dominant contributions to the mono-jet events come from initial state radiation of a jet. As a result, the events are predominantly distributed in the forward and backward regions. Fig.~\ref{fig:ca:pt:mj:LHC13} (a) and (b) show the normalized polar angle $(\theta_j)$ and transverse momentum $(p_{T,j})$ distributions of the jet in the laboratory frame with center of mass energy $\sqrt{s}=13\tev$, respectively. The signals induced by the operators $\calo_1$ and $\calo_2$ are shown by the cyan-dashed and blue-dashdotted curves for a massless dark fermion and energy scale $\varLambda_i = 1\tev$, and the irreducible background is shown by the black-solid curve.
One can clearly see the collinear and soft “singularity” behaviors in both the signal and background events, and the background events are softer than the signal events. Hence, the transverse momentum is a good observable to distinguish the signals and background. However, both the polar angle and transverse momentum distributions are exactly the same for the operators $\calo_1$ and $\calo_2$. Hence, the mono-jet events are unable to identify the Lorentz structures of the signal operators.
Fig.~\ref{fig:ca:pt:mj:LHC13} (c) shows the total parton-level cross sections of the signal and background as a function of the center-of-mass energy $\sqrt{\hat{s}}$. The signals are shown for a massless dark fermion and energy scale $\varLambda_i = 1\tev$. One can clearly see that the signal cross sections grow rapidly with increasing $\sqrt{\hat{s}}$, while the background cross sections decrease dramatically. Around $\sqrt{\hat{s}} \sim 1\tev$, the signal is roughly 5 orders of magnitude smaller than the background. The signal becomes comparable to the background only when $\sqrt{\hat{s}} \gtrsim 7\tev$. Consequently, we need a relatively strong cut on the transverse momentum $p_{T,j}$ to enhance the signal significance.

The ATLAS collaboration has searched for new phenomena in events containing an energetic jet and large missing transverse momentum \cite{ATLAS:2021kxv}. For an axial-vector mediated model, the exclusion limit for a massless dark fermion reaches about $2.1\tev$. It is expected that the four-fermion contact couplings can be constrained to a similar level. We use those data to estimate constraints on the parameters $m_\chi$ and $\varLambda_i$.
Validation of our simulation is given in the App.~\ref{app:valid:mj}. Since there is a strong cut on missing transverse energy ($p_{T, j} > 150$ GeV), our simulation is done at the generator level. 
The exclusion limits relevant to the LHC13 are ascertained via the following $\chi^2$ computation:
\bee
\chi^2  =  \sum_{i} 
\left[  \frac{ \epsilon_D \cdot N^{\rm S}_{i} }{ \sigma^{\rm ATLAS}_{i} } \right]^2\,.
\ene
In this context, $\sigma^{\rm ATLAS}_{i}$ stands for the experimental uncertainty associated with the $i$-th bin, as reported by the ATLAS research team, and $N^{\rm S}_{i}$ denotes the number of signal events in that bin. 
The total detector efficiency is accounted for by an overall normalization factor 
$\epsilon_D$
estimated through validating the irreducible background process $pp\to j Z(\nu\nu)$. 
Furthermore, there are also significant reducible contributions to the total background events. For example, the process $q\bar{q} \to W(\tau\nu)$ with subsequent leptonic decay of the $\tau$-lepton contributes to the reducible background.

For center-of-mass energy $\sqrt{s}=13\tev$, the total reducible contribution is about $40\%$ of the total mono-jet events \cite{ATLAS:2021kxv}. Since, in the signal region, the $p_{T,j}$ distribution of the reducible background exhibits a behavior similar to that of the irreducible background, their contributions to the total background are simply estimated by an overall scale factor in validating our simulation. This overall normalization factor approximation method is used to estimate detector-level predictions for both the (irreducible and reducible) background and signal processes.

\begin{figure}[h] 
\centering
\includegraphics[width=0.48\textwidth]{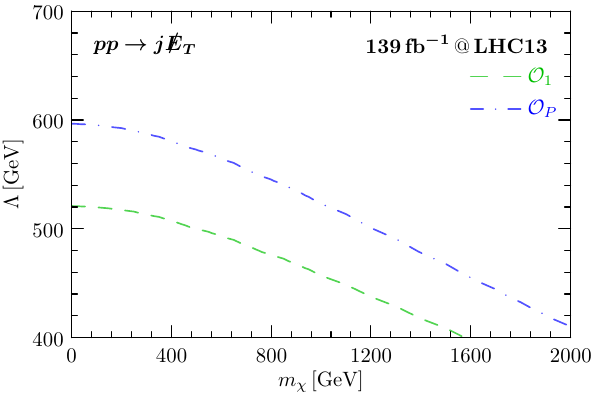}
\includegraphics[width=0.48\textwidth]{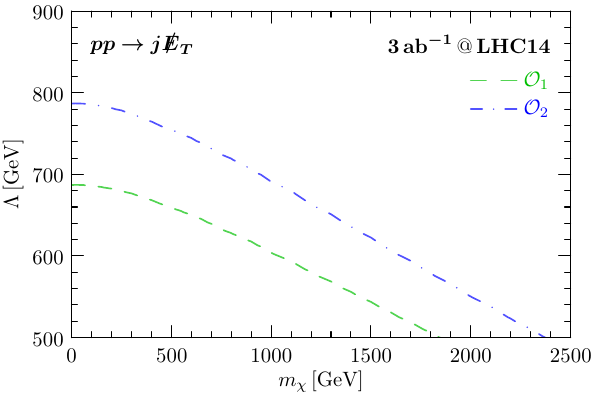}
\caption{\it 
Expected exclusion limits at 95\% C.L. obtained using the mono-jet production
process at the LHC with $\sqrt{s}=13\tev$
and a total luminosity $\call=139\fb^{-1}$ (\textbf{Right panel}),
and $\sqrt{s}=14\tev$ and a total luminosity $\call=3\ab^{-1}$ (\textbf{Left panel}).
}
\label{fig:mj:LHC1314}
\end{figure}

The left panel of Fig.~\ref{fig:mj:LHC1314} shows the 95\% expected exclusion limits in the $m_\chi - \varLambda$ plane for our signal operators at the LHC with center-of-mass energy $\sqrt{s}=13\tev$ and a total luminosity $\call=139\fb^{-1}$.
We can see that the strongest limit is given for the operator $\calo_2$. This is mainly due to the different scale factors in the definitions of the operators given in \eqref{eq:ggo:cpe} and \eqref{eq:ggo:cpo}: for the same event rate, the bound on the energy scale $\varLambda_2$ is always larger than the bound on $\varLambda_1$ by a factor of $(12/8)^{1/3} \approx 1.14$. For a massless dark fermion, i.e., $m_\chi \sim 0$, the bounds on the energy scales can reach about $600\gev$ and $520\gev$, respectively.
One can note that the constraints are much weaker than the typical collision energy, and hence the EFT description can be invalid in this region. However, the bounds are obtained by factorizing out an additional normalization factor: $\alpha_s/12\pi$ for the operator $\calo_1$ and $\alpha_s/8\pi$ for the operator $\calo_2$ (see \eqref{eq:ggo:cpe} and \eqref{eq:ggo:cpo}). Including these factors, the energy scales can reach about 3.5\,TeV for both operators. Considering this, the EFT description is valid in our case (at the hadron collider, other contributions with $\sqrt{\hat s} > 3.5\tev$ are automatically suppressed due to parton distribution functions).
The right panel of Fig.~\ref{fig:mj:LHC1314} shows the 95\% expected exclusion limits in the $m_\chi - \varLambda$ plane for center-of-mass energy $\sqrt{s}=14\tev$ and a total luminosity $\call=3\ab^{-1}$. One can see that the bounds are enhanced by a factor of about 1.3. This is not significant, and the reason is that the production rate is proportional to $1/\varLambda_{i}^{6}$.

At the LHC and future high-energy colliders, it is possible that the effective interactions occur at an energy scale higher than the cutoff, and hence unitarity can be violated \cite{Busoni:2013lha,Busoni:2014sya,Busoni:2014haa} (and references therein). Assuming that the momentum transfer $|Q|$ is smaller than the mass of the possible mediator $M$, i.e., $|Q|<M$, and the couplings are smaller than $4\pi$ for the validity of perturbation, it was shown that only those events with $|Q| < 4\pi \varLambda$ are reliable \cite{Busoni:2013lha,Busoni:2014sya,Busoni:2014haa}.
We have checked the influence of this cut on the exclusion limits and find it negligible. The main reason is that the cut $|Q| < 4\pi \varLambda$ is not stringent (for instance, for the operator $\calo_{1}$ at LHC13 (see the left panel of Fig.~\ref{fig:mj:LHC1314}), the cut is $|Q| < 6.5$ TeV), and events in this region are inherently suppressed by the PDF of the protons \cite{Ellis:2021dfa}. However, a large fraction of the events can be removed if a stringent cut, say $|Q| < \varLambda$, is applied \cite{Busoni:2014sya}. In this case, our exclusion limits can be significantly reduced.
The cutoff scale cannot be precisely defined unless the UV completion of the EFT operators is given. For a model-independent approach, we recommend the most conservative cutoff $|Q| < 4\pi \varLambda$ allowed by perturbativity. On the other hand, a cut $\varLambda > m_{\chi}/(2\pi)$ is also used as a benchmark for the validation of the EFT. 
In this work, the dark fermion $\chi$ is interpreted as a possible DM candidate only when its mass is $\sim (1, 100)$ MeV (see Sec.~\ref{sec:AbsNT}). In this range, it is clear that the condition  $\varLambda > m_{\chi}/(2\pi)$ is satisfied.

\begin{figure}[h]
\centering
\includegraphics[width=0.32\textwidth]{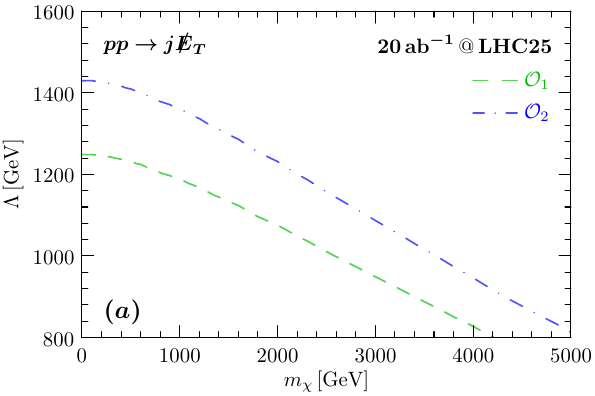}
\includegraphics[width=0.32\textwidth]{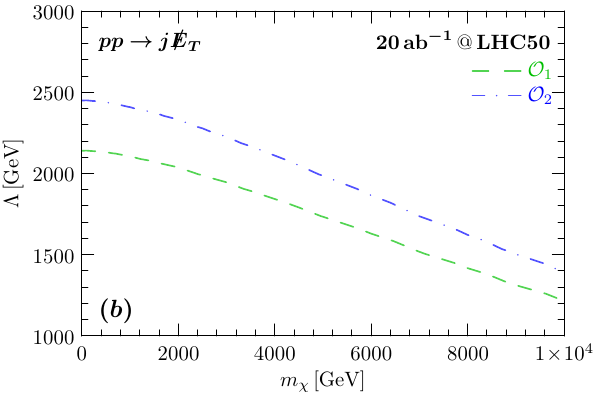}
\includegraphics[width=0.32\textwidth]{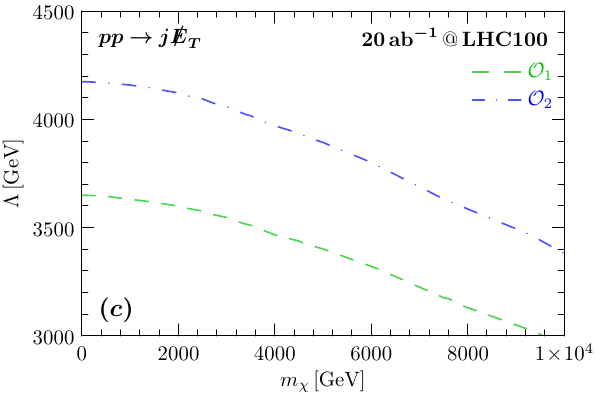}
\caption{\it 
Expected exclusion limits at 95\% C.L. obtained using the mono-jet production
process at the LHC with an universal luminosity
$\call=20\ab^{-1}$ and center of mass energy 
$\sqrt{s}=25\tev$ \textbf{(a)}, $50\tev$ \textbf{(b)} and $100\tev$ \textbf{(c)}, respectively.
}
\label{fig:exclusion:mj:HHC}
\end{figure}

In addition to the current LHC and its upgrade to HL-LHC, there are also other proposed hadronic colliders, including HE-LHC \cite{CidVidal:2018eel}, FCC-hh \cite{FCC:2018vvp}, and SppC \cite{CEPC-SPPCStudyGroup:2015csa}, whose collision energies range from 25\,TeV to 50\,TeV and 100\,TeV. We extend our study to those cases. Fig.~\ref{fig:exclusion:mj:HHC} (a), (b) and (c) show the results for the cases with a universal luminosity $\call=20\ab^{-1}$ and center-of-mass energy $\sqrt{s}=25\tev$, $50\tev$ and $100\tev$, respectively. 
One can see that constraints on the energy scales start to exceed 1\,TeV for a massless dark fermion. If $\sqrt{s}=50\tev$, for the signal operators with $\varLambda_i = 1\tev$, a dark fermion with a mass of about $2\sim 3 \tev$ can be excluded.

\section{Absorption at Nuclear Targets}
\label{sec:AbsNT}
It is well-known that the parton-level operators can induce non-trivial interactions at the nucleon level, which can be investigated by dark fermion scattering off a nuclear target \cite{Goodman:1984dc,Cebrian:2021mvb,DiGangion:2021thw,PICO:2022nyi,Bishara:2017pfq}. For instance, the four-fermion contact interactions have been studied in Refs.~\cite{Dror:2019onn,Dror:2019dib,Dror:2020czw,Ge:2022ius,Li:2022kca,Ge:2023wye,Ma:2024aoc,Ge:2024euk,PandaX:2022osq,PandaX:2022ood,CDEX:2024bum}.
Here we focus on the gluonic interactions defined in \eqref{eq:ggo:cpe} and \eqref{eq:ggo:cpo}. There are different conventions in the parameterization of the nucleon-level matrix elements \cite{Belanger:2008sj,Cirelli:2013ufw,Bishara:2017pfq,Crivellin:2013ipa,Shifman:1978zn,Drees:1993bu}. In this work, we employ the parameterization used in Refs.~\cite{Bishara:2017pfq,Crivellin:2013ipa,Shifman:1978zn,Drees:1993bu}, and for completeness, we also list them below:
\bea
\label{eq:nm:cpe}
\left\langle N^{\prime}\left|\frac{\alpha_s}{12 \pi} G^{a \mu \nu} G_{\mu \nu}^a\right| N\right\rangle 
& =
F_G^N\left(q^2\right) \bar{u}_N^{\prime} u_N \,, 
\\[3mm]
\label{eq:nm:cpo}
\left\langle N^{\prime}\left|\frac{\alpha_s}{8 \pi} G^{a \mu \nu} \widetilde{G}_{\mu \nu}^a\right| N\right\rangle 
& =
F_{\widetilde{G}}^N\left(q^2\right) \bar{u}_N^{\prime} i \gamma_5 u_N \,,
\ena
where $F_G^N\left(q^2\right)$ and $F_{\widetilde{G}}^N(q^2)$ are nucleon form factors and usually are functions of the squared momentum transfer $q^2 = (p_f - p_i)^2 = (p_\nu - p_\chi)^2$. In the approximation that both the DM and the initial nucleus are at rest, the squared momentum transfer is simply related to the mass of the DM as $q^2 \approx - m_\chi^2$. Since we are interested in the DM with mass parameter $m_\chi \lesssim m_N$, the approximation $q^2\sim 0$ is usually adopted for the evaluations of the form factors. Details of the form factors can be found in App.~\ref{sec:App:ff}.

Hadron-level interactions can induce either elastic or inelastic scattering of the DM off a nuclear target, depending on the model of the dark sector \cite{Hisano:2017jmz,Misiaszek:2023sxe,Schumann:2019eaa}. In our case, since the dark fermion is completely converted to a neutrino, the scattering is inherently inelastic and is also known as the absorption process \cite{Dror:2019onn,Dror:2019dib}.
More specifically, the matrix element given in \eqref{eq:nm:cpe} and \eqref{eq:nm:cpo} can be probed by searching for the recoil energy of the final nucleus state of the following inelastic scattering:
\bee
\label{eq:abs:proc:nuc}
\chi(p_\chi) + A(p_i) \rightarrow \nu(p_\nu) + A(p_f) \,,
\ene
where the symbol $A$ (the mass number of the isotope under consideration) is used to specify the nucleus target, and the momenta of the initial and final states are explicitly specified.
The differential scattering rate per nuclear recoil energy is given as:
\bee
\label{eq:dd:drt}
\frac{\d R_A}{\d E_R}
= 
N_A n_\chi
\int \mathrm{d}^3 v f_{\mathrm{E}}(\boldsymbol{v}_\chi, t) v_\chi 
\frac{\mathrm{d} \sigma_A}{\mathrm{~d} E_{\mathrm{R}}} \,, 
\ene
where $N_A$ is the number of nuclei of the target (if there is more than one kind of isotope/nucleus, the total rate is simply the sum of the individual contributions); $n_\chi =  \rho_\chi/m_\chi$ is the local number density of the DM with the local energy density $\rho_\chi \simeq 0.3 {\,\rm GeV/cm}^3$; $\d\sigma_{\rm A}/\d E_{\rm R}$ is the differential cross section of the absorption process defined in \eqref{eq:abs:proc:nuc}, and $f_{\mathrm{E}}(\boldsymbol{v}_\chi, t)$ is the velocity distribution of the incoming DM.

Since the DM is assumed to be non-relativistic, i.e., $m_\chi \gg p_\chi$, the dominant contribution to the amplitude is given in the limit $p_\chi \to 0$. As a result, the leading-order amplitude of the absorption process ($\mathcal{M}_A$) is independent of the DM velocity. In this case, the velocity integral in \eqref{eq:dd:drt} can be carried out independently.
Furthermore, the leading-order recoil energy of the nucleus is proportional to the squared mass of the DM, $E_{R}^0 = m_\chi^2/2m_{A}$, where $m_{A}$ is the mass of the nucleus. Consequently, the differential cross section $\d\sigma_{\rm A}/\d E_{\rm R}$ exhibits a sharp peak at $E_{R}^0$.
In addition, a step function $\varTheta(E_{\rm R} - E_{\rm R}^{\rm th})$, with $E_{\rm R}^{\rm th}$ being the threshold of the recoil energy, is employed in practice to account for the fact that there is a minimal detectable energy in a specific experiment.
Incorporating the phase space and flux factors, and carrying out the velocity integral, given that the dark matter (DM) velocity distribution function must be normalized to unity, one can readily determine that the differential scattering rate can be expressed as
\bee
\frac{\mathrm{d} R_A}{\mathrm{d} E_{\mathrm{R}}}
=
\frac{ N_A  n_\chi }{16 \pi m_{A}^2  } \overline{\big| \calm_A \big|^2}
\delta\big( E_R - E_R^0 \big)  \varTheta(E_{\rm R} - E_{\rm R}^{\rm th}) \,.
\ene
Here, $\overline{ | \mathcal{M}_A |^2 }$ represents the squared amplitude of the process described in Eq. \eqref{eq:abs:proc:nuc}, where the helicities of the initial and final states are averaged and summed respectively.
For a nucleus having total spin $J$, the average of the squared amplitude $\overline{ | \mathcal{M}_A |^2 }$ is explicitly defined as:
\bee
\overline{|\mathcal{M}_A|^2} 
\equiv 
\frac{1}{2 s_{\chi}+1} \frac{1}{2 J+1} 
\sum_{\substack{\text { initial \& } \\ \text { final spins }}}|\mathcal{M}_A|^2 \,,
\ene
where $s_{\chi}=1/2$ is the spin of the DM particle. For a DM with mass $m_\chi \lesssim 100\mev$, its de Broglie wavelength is comparable to the size of a typical nucleus. The interactions between the DM and the nucleus are coherent.
Usually, the scatterings are classified as spin-independent (SI) and spin-dependent (SD) processes to account for the destructive coherence of the spin states of the nucleons inside the nucleus. The SI and SD interactions are distinguished by a coherence factor, and the corresponding nuclear response functions are completely different and experiment-dependent. We will discuss the details later.

From \eqref{eq:nm:cpe} and \eqref{eq:nm:cpo}, one can see that the CP-even and CP-odd gluonic operators can induce scalar and pseudo-scalar interactions at the nucleon level, respectively. It is well known that the leading-order non-relativistic expansions of the scalar and pseudo-scalar operators of the nucleon fields can induce SI and SD scatterings \cite{Ma:2024tkt}, respectively. Hence, these two operators have different responses in the scattering experiment, and we will study them separately.

\subsection{Spin-Independent Absorption}
\label{sec:DD:SI}
As shown in \eqref{eq:nm:cpe}, the CP-even gluonic operator at low energy is matched to the scalar interaction among nucleons. In the non-relativistic limit, the scattering amplitude is given as:
\bee
\label{eq:nro:ii}
\calm_{G}^{N}
\approx
2\sqrt{2} m_N m_\chi  F_{G}^{N}
\big[(\xi_{h_\nu}^{\nu})^\dag \, \xi^{\chi}_{h_\chi} \big]
\big[(\omega_{h_{N'}}^{N'})^\dag \,\omega_{h_N}^N \big] \,,
\ene
where $\xi_{h_a}^a$ are the 2-component spinors with helicity $h_a$ for the neutrino ($a=\nu$) and the DM ($a=\chi$), and $\omega_{h_a}^a$ are the 2-component spinors with helicity $h_a$ for the incoming nucleon ($a=N$) and the outgoing nucleon ($a=N'$). 
Evidently, the above matrix element is independent of the nuclear spin. Consequently, only SI scattering contributes to the absorption process.
In this case, the amplitude at the nuclear level is straightforwardly related to the amplitude at the nucleon level as shown below:
\bee
\calm_A = \sum_{N=p, n} F_{\rm Helm}(q^2)  C_N \calm_{N}  \,.
\ene
Here, $F_{\rm Helm}(q^2)$ represents the response function of the nuclei for SI scattering and is recognized as the Helm form factor \cite{Helm:1956zz, Lewin:1995rx},
and $C_N$ represents the coherence factor. In SIprocesses, $C_N$ takes the value of $Z$ for protons and $A-Z$ for neutrons. Conversely, in SD interactions, the coherence factor simplifies to $C_N = 1$, suggesting that protons and neutrons contribute uniformly without interference. 
Here we use the formulation suggested in Refs.~\cite{Engel:1992bf,Jungman:1995df,Engel:1991wq}.
Furthermore, it is well-known that isospin symmetry conservation is an excellent approximation in the DM scattering experiment. With the help of this approximation, the difference of the amplitudes of scattering off a proton and a neutron is negligible, i.e., $\calm_p \approx \calm_n$. Then the nuclear level scattering amplitude is simply given as $\calm_A^{S} = F_{\rm Helm}(q^2) A \,\calm_{N}^{S}$.
With straightforward calculations, the squared average of the amplitude given in \eqref{eq:nro:ii} can be easily obtained:
\bee
\overline{|\mathcal{M}_{G}^{N}|^2}
=
\frac{4 m_{A}^2 m_\chi^2 }{ \varLambda^6_1 } \big(F_{G}^{N} \big)^2\,.
\ene
where we have assumed that the DM couples to quarks universally.

Collecting all the terms, we can find the scattering rate as follows:
\bee
R_G^{A}
=
N_A  n_\chi  \frac{ m_{\chi}^2 }{4 \pi  \varLambda_{1}^6 } 
\big[ A F_{G}^{N}(m_\chi^2) F_{\rm Helm}(m_\chi^2) \big]^2
\varTheta(E_{\rm R}^0 - E_{\rm R}^{\rm th}) \,,
\ene
where we have used the approximation $q^2 \approx m_\chi^2$.

\begin{figure}[h]
\centering
\includegraphics[width=0.79\textwidth]{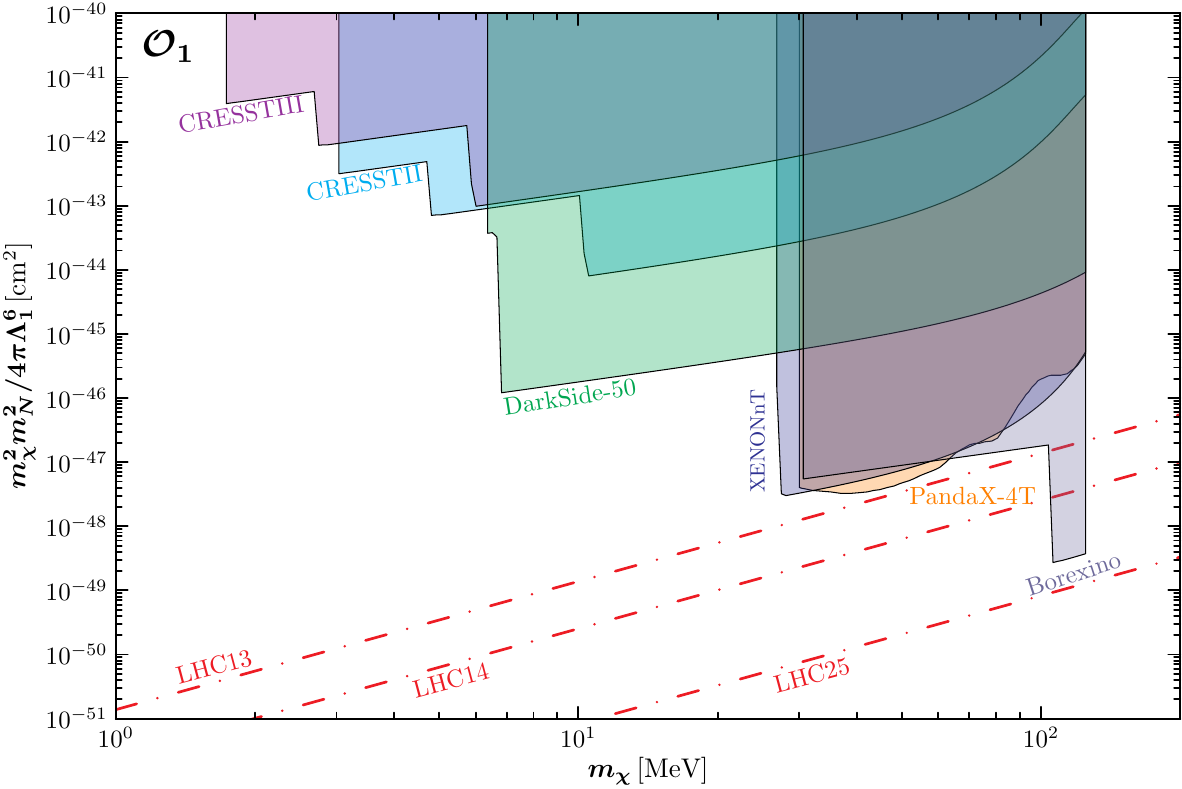}
\caption{\it 
Excluded regions in the $m_\chi$-$\sigma_{gg}$ plan
by SI absorption of the DM particle at nuclear targets.
The SI interactions are induced by the operator $\calo_1$.
}
\label{fig:SI:CS}
\end{figure}

The PandaX collaboration has searched for absorption of a DM at a nuclear target \cite{PandaX:2022osq}, and the limit on the SI scattering cross section for a fermionic DM with mass $40\mev/c^2$ is $1.5 \times 10^{-50} {\rm cm^2}$ at 90\% $\rm C.L$. Here we reinterpret the results as constraints for the operator $\calo_1$.
We will investigate the other experiments listed in Tab. \ref{tab:SI:Exposure}, and the corresponding bounds are obtained by requiring that the total number of events is greater than 10. Fig. \ref{fig:SI:CS} shows the excluded regions in the $m_\chi$-$\sigma_{gg}$ plane, where the cross section $\sigma_{gg}$ is defined as:
\bee
\sigma_{gg} = \frac{m_\chi^2m_N^2}{4\pi\varLambda_1^6} \,,
\ene
with $m_N=(m_p + m_n)/2$ being the averaged mass of the proton and neutron.
The kinks are results of the change of the recoil energy for a particular isotope. 
One can clearly see that almost all the regions accessible by the direct detection experiments have been excluded by the constraints from the LHC with center of mass energy $\sqrt{s}=13\tev$ and a total luminosity $139\fb^{-1}$. The only exception is the DM absorption by the $^{*}_{6}\mathrm{C}$ isotopes in the Borexino experiment. The reason is simply the coherence enhancement effect of the SI scattering.
For the Borexino experiment, the absorption cross section of $^{*}_{6}\mathrm{C}$ is about two orders of magnitude larger than that of hydrogen.

\subsection{Spin-Dependent Absorption}
\label{sec:DD:SD}

From the nucleon matrix element in \eqref{eq:nm:cpo}, one can clearly see that the CP-odd gluonic operator at low energy can induce a pseudo-scalar interaction among nucleons, which results in a SD coupling in the non-relativistic limit.
Including the DM-neutrino sector, the leading-order non-relativistic expansion of the scattering amplitude is given as
\bee
\label{eq:nro:ss}
\calm_{\widetilde{G}}^{N}
=
2\sqrt{2} m_\chi  F_{\widetilde{G}}^{N}
\big[h_\nu(\xi_{h_\nu}^{\nu})^\dag \xi^{\chi}_{h_\chi} \big] \cdot
\big[(\omega_{h_{N'}}^{N'})^\dag  \big( \boldsymbol{q} \cdot \boldsymbol{s}\, \big) \,\omega_{h_N}^N \big] \,,
\ene
Since the absorption process is coherent, the total amplitude is the coherent sum of the individual nucleons. For a nucleus state with total spin $J$ and helicity $M$, the total amplitude is given as
\bee
\label{eq:cpo:ma}
\calm_{\widetilde{G}}^{A}
=
2\sqrt{2} m_\chi \sum_{N=p, n} 
F_{\widetilde{G}}^{N}
h_\nu \left\langle s_\nu, h_\nu \right. \left|  s_\chi, h_\chi \right\rangle \;
\big[ \left\langle J, M^{\prime}\left|
\boldsymbol{S}_N \cdot  \boldsymbol{q}
\right| J, M\right\rangle  \big] \,,
\ene
where $\boldsymbol{S}_N$ is the total nucleon spin operator defined as $\boldsymbol{S}_p \equiv \sum_{i=\text{protons}} \boldsymbol{s}_{p_i}$ and $\boldsymbol{S}_n \equiv \sum_{i=\text{neutrons}} \boldsymbol{s}_{n_i}$ for protons and neutrons, respectively.
For a spin-less nuclear state ($J = 0$), the above amplitude vanishes. Hence, non-trivial SD signals can only be seen in experiments that contain isotopes with non-zero total spin. In this paper, we study the SD signals at some typical experiments listed in Tab. \ref{tab:SD:Exposure}.
The squared amplitudes with summing over the final-state spins and averaging over the initial states are given as (there is an additional factor $1/2$ because only left-handed neutrinos can participate in the scattering)
\bee
\label{eq:ls:sm}
\overline{\big| \calm_{\widetilde{G}}^{A} \big|^2}
=
2 m_\chi^2 \sum_{N, N'} F_{\widetilde{G}}^{N} F_{\widetilde{G}}^{N'}  F_{10,10}^{NN'} 
\ene
and the response functions are defined as
\bee
\label{eq:ls:sm2}
 F_{10,10}^{NN'} 
=
\frac{1}{2J + 1}
\sum_{M, M'} 
\left\langle J, M\left|\boldsymbol{S}_{N'} \cdot \boldsymbol{q} \right| J, M'\right\rangle
\left\langle J, M'\left|\boldsymbol{S}_N \cdot \boldsymbol{q} \right| J, M\right\rangle 
 =
 \frac{q^2}{4} F_{\Sigma''}^{NN'}(q^2)  \,.
\ene
There is only the $\Sigma^{''}$ nuclear response function \cite{DelNobile:2021wmp} related to the amplitude given in \eqref{eq:cpo:ma}. In general, the response function $F_{\Sigma''}^{NN'}(q^2)$ is a complex function of the momentum transfer and model parameters. Explicit expressions of $F_{\Sigma''}^{NN'}$ for some isotopes can be found in Ref. \cite{Fitzpatrick:2012ix}. Here, we simply employ the zero momentum transfer approximation.
With the help of the following relation at zero momentum transfer
\bee
\label{eq:ssr}
 \sum_{M, M'} 
\left\langle J, M\left|S_{N', i}\right| J, M'\right\rangle
\left\langle J, M'\left|S_{N, j}\right| J, M\right\rangle
=
\frac{(J + 1)(2 J + 1)}{3 J} 
\mathbb{S}_N \mathbb{S}_{N'} \delta_{i j} \,,
\ene
where $\mathbb{S}_N \equiv \left\langle J, J \left|S_N^z\right| J, J\right\rangle$ are the expectation values of the nucleus spin operator for the states of maximal angular momentum, the response functions are given as
\bee
\label{eq:ss:boundary}
F_{10,10}^{NN'} 
=
\frac{J + 1}{3J} q^2 \mathbb{S}_N \mathbb{S}_{N'}  \,,
\quad\quad
F_{\Sigma''}^{NN'} = \frac{4(J + 1)}{3J}  \mathbb{S}_N \mathbb{S}_{N'} \,.
\ene
Here, the $q^2$ dependence of the response function $F_{10,10}^{NN'}$ is factorized out, and we will use the approximation $q^2 \approx m_\chi^2$ in the following calculations.
The value of $\mathbb{S}_N$ is somewhat model-dependent since its determination requires detailed calculations within realistic nuclear models. In Tab.~\ref{tab:exps}, we list $\mathbb{S}_N$ of some isotopes studied in this paper. We refer to Ref. \cite{DelNobile:2021wmp} for more details of $\mathbb{S}_N$.
The scattering rate is usually calculated by assuming that either protons or neutrons participate in the interaction. Here, we estimate the scattering rate by assuming that only the nucleon with the largest spin expectation value can contribute. Including the phase space factor, the scattering rate is given as
\bee
\label{eq:SD:rate:p}
R_{\widetilde{G}}^{A}
=
\frac{ N_A  n_\chi  m_{\chi}^4 }{8 \pi  \varLambda_{2}^6 m_{A}^2 } 
\varTheta(E_{\rm R}^0 - E_{\rm R}^{\rm th})
\sum_{N,N'=p, n} F_{\widetilde{G}}^{N} F_{\widetilde{G}}^{N'} 
\big[ F_{\Sigma''}^{NN'} \big]  \,.
\ene

\begin{figure}[h]
\centering
\includegraphics[width=0.79\textwidth]{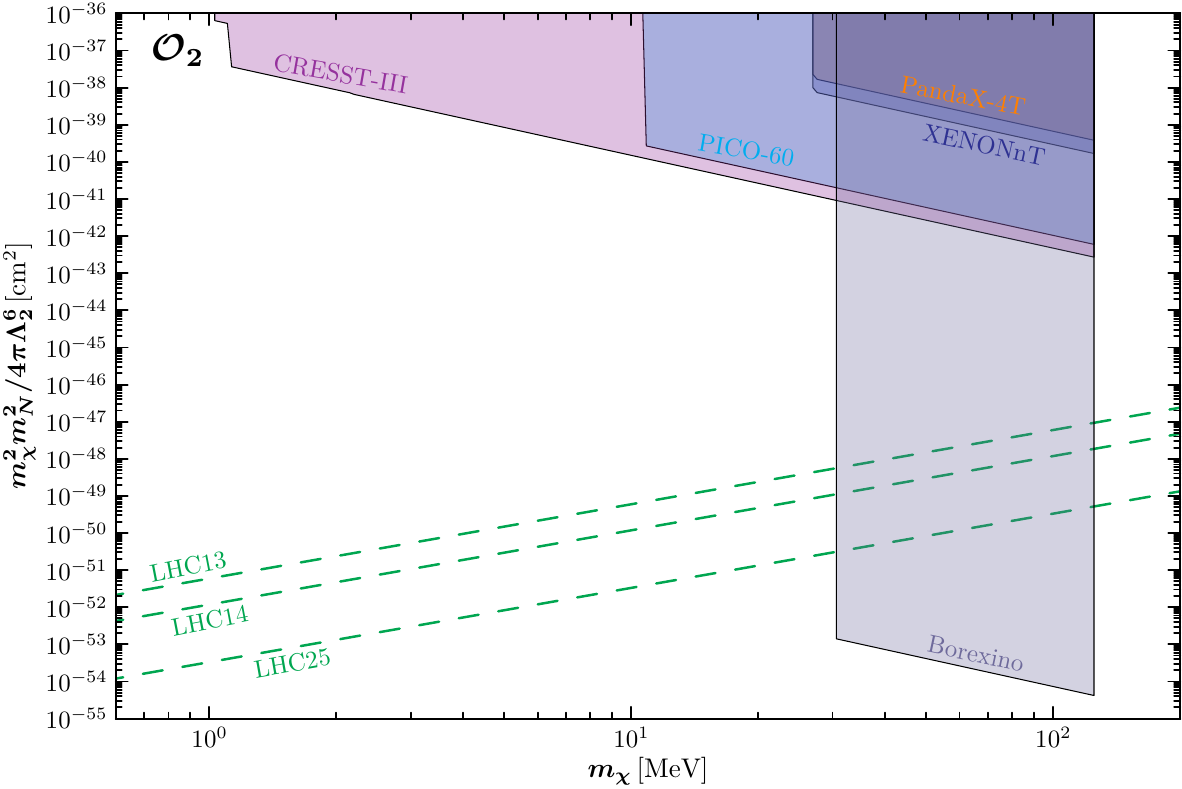}
\caption{\it 
Excluded regions in the $m_\chi$-$\sigma_{gg}$ plan
by SD absorption of the DM particle at nuclear targets.
The SD interactions are induced by the operator $\calo_2$.
}
\label{fig:SD:CS}
\end{figure}

Fig. \ref{fig:SD:CS} shows the expected excluded regions in the $m_\chi$-$\sigma_{gg}$ plane for the experiments listed in Tab. \ref{tab:SD:Exposure}. The exclusion lines are obtained by requiring that the total number of events is greater than 10.
Similar to the SI scattering case, one can clearly see that, except for the Borexino experiment, the parameter regions accessible by the other experiments are already excluded by the constraints from the LHC with center of mass energy $\sqrt{s}=13\tev$ and a total luminosity $139\fb^{-1}$. The reasons are as follows: firstly, the Borexino experiment has roughly 3 orders of magnitude larger exposure; secondly, for the Borexino experiment, the most significant contribution to the SD absorption comes from hydrogen. As a result, the number of nuclei is more than 2 orders of magnitude larger than that in the PandaX or XENON experiment; thirdly, compared to the SI scattering, the SD scattering rate is suppressed by a factor of $m_\chi^2/m_A^2$. Hence, the scattering rate of the Borexino experiment is about $10^4$ larger than that of the PandaX/XENON experiment; finally, the form factor $F_{\widetilde{G}}^{N} \sim m_N$ is also more than 1 order of magnitude larger than the form factor $F_{G}^{N} \sim 2 m_G/27$.
In total, the constraints from the Borexino experiment are roughly 11 orders of magnitude stronger than the bounds from the other experiments, as one can see in Fig. \ref{fig:SD:CS}. 
Therefore, the direct detection experiment with light isotopes is much more sensitive to the SD absorption. This is also true for the SD absorption induced by operators involving quark pairs \cite{Ma:2024tkt}.

\section{Conclusion}
\label{sec:conclusion}

In this paper, we study constraints on the gluonic contact interaction operators involving a dark fermion and a SM neutrino, which are explicitly defined in \eqref{eq:ggo:cpe} and \eqref{eq:ggo:cpo}. The invisibility of the dark fermion at high-energy colliders is guaranteed in the mass range $m_\chi \lesssim 20\gev$, as shown in Fig.~\ref{fig:dcylen}. For a heavier dark fermion, a more complete model is necessary, but it will introduce more parameters at the same time. Here, we employ the EFT approach to model-independently study constraints on the signal operators at hadron colliders and direct detection experiments. The advantage of this approach is that one can easily convert our results to bounds on the parameters of a UV-completed model. Furthermore, one of the promising features of high-energy colliders is that very massive particles can be searched for as long as it is kinematically allowed. Hence, it is reasonable to study signals of the dark fermion at hadron colliders in the full mass range.

At the LHC, we show that mono-jet production is a promising probe of the gluonic contact operators. By using current LHC results at a center-of-mass energy $\sqrt{s}=13\tev$, we obtained lower bounds on the scale parameters of about 520 GeV and 600 GeV for the operators $\calo_{1}$ and $\calo_{2}$, respectively. The bounds can be slightly improved at the HL-LHC ($\sqrt{s}=14\tev$ and a total luminosity $\call = 3\ab^{-1}$), and can reach about 1.22 TeV and 1.42 TeV at the HE-LHC ($\sqrt{s}=25\tev$ and a total luminosity $\call = 20\ab^{-1}$). These bounds can be further improved at hadron colliders with $\sqrt{s}=50\tev$ and $\sqrt{s}=100\tev$.

We also studied the signal properties of the gluonic contact operators at direct detection experiments. Since the dark fermion can be converted to a neutrino at a nuclear target, the signal is represented by the nuclear recoil energy after absorption of the dark fermion. We show that while the operator $\calo_{1}$ can induce spin-independent absorption, the interactions induced by the operator $\calo_{2}$ are spin-dependent. We studied the experimental sensitivities for some typical direct detection experiments, and the expected constraints are obtained by requiring that the total number of scattering events is less than 10.

We find that almost all the parameter space accessible by the spin-independent absorption has been excluded by the current LHC constraints. In contrast, for spin-dependent absorption at a light nuclear target, there is still some parameter space ($m_\chi$ in the range $\sim [10, 100]\mev$) that cannot be reached by current and upcoming LHC searches. As we know, so far there is only an absorption experiment with a Xenon target performed by the PandaX collaboration \cite{PandaX:2022osq}. Hence, this is particularly interesting for probing the dark fermion at direct detection experiments with a light nuclear target, such as the Borexino experiment. Apart from this range, collider searches are much more promising and can provide a complementary approach to looking for a neutrino-philic dark fermion.

\appendix

\section{Validating Simulation of the Mono-jet Events}
\label{app:valid:mj}

\begin{figure}[h] 
\centering
\includegraphics[width=0.68\textwidth]{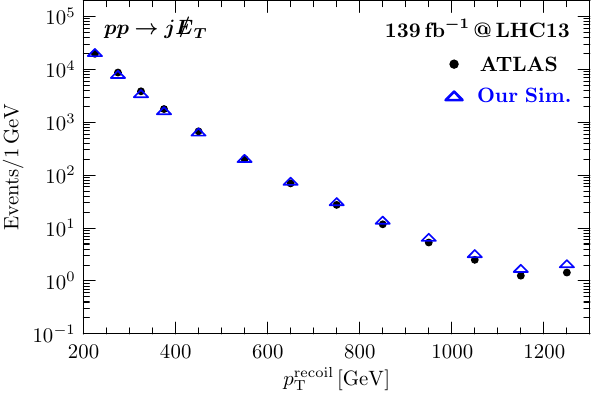}
\caption{\it 
Validation of our simulation for the $p_{T}^{\rm recoil}$ 
distribution of the irreducible background
channel $pp\to j Z(\nu\bar\nu)$ at the LHC with a total luminosity $\call = 139\fb^{-1}$.
The experimental data (black dots) are taken from the Ref. \cite{ATLAS:2021kxv},
and our results (red rectangles) have been renormalized by multiplying an overall constant.
}
\label{fig:valid:mj:LHC13}
\end{figure}

The ATLAS collaboration has searched for new phenomena in events containing 
an energetic jet and large missing transverse momentum \cite{ATLAS:2021kxv}.
The events are selected according to the kinematic cuts 
$E_{T}^{\rm miss} > 200\gev$, $p_{T, j} > 150$ GeV and $|\eta_j| < 2.4$
 \cite{ATLAS:2021kxv}. 
In this region the parton shower effect is negligible due to 
strong cut on missing transverse energy.
In consideration of this, our simulation is done at the generator level.
The total detector efficiency is estimated by an overall 
normalization factor which is obtained by validating the irreducible 
background process $pp\to j Z(\nu\nu)$. 
The same normalization factor is also used to estimate detector level prediction 
for both the background and signal. 
The Fig.~\ref{fig:valid:mj:LHC13} shows comparison between 
our simulation (red square) and the ATLAS result (black dot)
for the $p_{T}^{\rm recoil}$ (which is equivalent to the transverse missing energy 
or transverse momentum of the jet at the generator level)
distribution of the irreducible background
channel $pp\to j Z(\nu\bar\nu)$ at the LHC13 with a total luminosity $\call = 139\fb^{-1}$.
One can see that in the signal region the parton shower and detector effects 
can be well described.  
It is clear that the approximation of an overall 
normalization factor works well not only for the total number of events 
but also the differential distributions. 
Hence, within the experimental uncertainty our simulation is valid.
This is particularly important 
in estimation of the exclusion limit which is obtained by calculating 
$\chi^2$ of the $p_{T}^{\rm recoil}$ distribution.

\section{Form Factors of the Nucleon Matrix Elements}
\label{sec:App:ff}

In this paper we consider only the leading order contributions of the form factors.
In the limit of zero momentum transfer, the form factors are given as
\cite{Bishara:2017pfq,Bishara:2016hek,Gross:1979ur},
\bea
F_G^N(0) &=& -\frac{2 m_G}{27} \,,
\\[3mm]
F_{\widetilde{G}}^N(0) 
&=&
-\widetilde{m} m_N\left[\frac{\Delta u}{m_u}+\frac{\Delta d}{m_d}+\frac{\Delta s}{m_s}\right] \,,
\ena
Here the nonperturbative coefficient $m_G$ accounts for the gluonic contribution 
to the nucleon mass in the isospin limit. The specific numerical values and relation  of these coefficients at the energy scale $Q=2 \mathrm{GeV}$ are
given as,
\bea
&&
\widetilde{m}^{-1} = \left( m_u^{-1} + m_d^{-1} + m_s^{-1} \right)\,, 
\\[3mm]
&& \Delta u=0.897(27) \,, \quad \Delta d=-0.376(27) \,, \quad \Delta s=-0.031(5)\,,
\\[3mm]
&&
 m_G = 848\pm 14 \mev\,, \quad
m_u = 2.14 \pm 8 \mev\,, 
\\[3mm]
&&
m_d = 4.70 \pm 5 \mev\,,  \quad
m_{s} = 94.05 \pm 1.03\mev\,.~~
\ena

\section{Experimental Parameters for SI and SD Absorption}
The Tab. \ref{tab:SI:Exposure} lists the experiments
studied in this paper for probing the SI absorption signals.
Some typical experiments, 
which contains parts of isotopes having non-zero total spin,
are listed in the Tab. \ref{tab:SD:Exposure}.
We will study the SD signals at these experiments.
In the Tab.~\ref{tab:exps}, we summarize the spin expectation values 
($\mathbb{S}_N$) of some isotopes that will be studied in this paper.
\begin{table}[th]
\renewcommand\arraystretch{1.58}
\begin{center}
\begin{tabular}{c c c c c}
~~~Experiment~~~   & ~~~~~~~~Taget~~~~~~~~ & ~~~~~Exposure~~~~~
& ~~~~~$E_{\rm R}^{\rm th}$~~~~~  
\\\hline\hline
CRESSTII \cite{CRESST:2015txj}
&
$\mathrm{Ca}\mathrm{W}\mathrm{O}_4$
& 
$52$ kg day
&
307 eV
\\[2mm]\hline
CRESSTIII \cite{CRESST:2017cdd}
&
$\mathrm{Ca}\mathrm{W}\mathrm{O}_4$
& 
$2.39$ kg day
&
100 eV
\\[2mm]\hline
DarkSide-50 \cite{DarkSide:2014llq,DarkSide:2018bpj}
&
Liquid $\mathrm{Ar}$
& 
$6787$ kg day
&
0.6 keV
\\[2mm]\hline
XENONnT \cite{XENON:2023cxc} 
& 
Liquid Xe
& 
1.09 t yr 
&
3 keV
\\[2mm]\hline
PandaX-4T \cite{PandaX:2023xgl} 
& 
Liquid Xe 
& 
0.55 t yr
&
3 keV
\\[2mm]\hline
Borexino \cite{Borexino:2018pev}
& 
$\mathrm{C}_6 \mathrm{H}_3\left(\mathrm{CH}_3\right)_3$
& 
817 t yr
&
500 keV
%
%
\\[2mm]
\hline\hline
\end{tabular}
\caption{\it 
Experiments studied here for probing the SI absorption signals.}
\label{tab:SI:Exposure}
\end{center}
\end{table}
\begin{table}[th]
\renewcommand\arraystretch{1.58}
\begin{center}
\begin{tabular}{c c c c c c}
Experiment  & Target & Exposure & Isotope (Abund.) & $E_{\rm R}^{\rm th}$
\\\hline\hline
XENONnT \cite{XENON:2023cxc}
& 
Liquid Xe 
& 
1.09 t yr 
&
\begin{tabular}{c} 
$ {}^{129}_{\;\;54}\mathrm{Xe} $ (26.4\%) \\ 
$ {}^{131}_{\;\;54}\mathrm{Xe} $ (21.2\%) 
\end{tabular}
& 
3 keV
\\[2mm]\hline
PandaX-4T \cite{PandaX:2022xas}
&
Liquid Xe
& 
0.63 t yr 
& 
\begin{tabular}{c} 
$ {}^{129}_{\;\;54}\mathrm{Xe} $ (26.4\%) \\ 
$ {}^{131}_{\;\;54}\mathrm{Xe} $ (21.2\%) 
\end{tabular}
& 
3 keV
\\[2mm]\hline
Borexino \cite{Borexino:2018pev}
& 
$\mathrm{C}_6 \mathrm{H}_3\left(\mathrm{CH}_3\right)_3$
& 
817 t yr
&
\begin{tabular}{c} 
$ {}^{13}_{\;\,6}\mathrm{C} $ (1.1\%) \\
$ {}^{1}_{1}\mathrm{H} $ (99.985\%) 
\end{tabular}
&
500 keV
\\[2mm]\hline
CRESST-III \cite{CRESST:2022dtl}
&
$\mathrm{Li}\mathrm{Al}\mathrm{O}_2$
& 
2.345 kg day 
&
\begin{tabular}{c} 
$ {}^{6}_{3}\mathrm{Li} $ (7.5\%) \\ $ {}^{7}_{3}\mathrm{Li} $ (92.5\%)
\\ $ {}^{27}_{13}\mathrm{Al} $ (100\%) 
\end{tabular}
& 
94.1 eV
\\[2mm]\hline
PICO-60 \cite{PICO:2019vsc}
& 
$\mathrm{C}_3 \mathrm{F}_8$
& 
2207 kg day 
&
\begin{tabular}{c} 
$ {}^{13}_{\;\,6}\mathrm{C} $ (1.1\%) \\
$ {}^{19}_{\;\,9}\mathrm{F} $ (100\%) 
\end{tabular}
&
3.3 keV
%
%
\\[2mm]
\hline\hline
\end{tabular}
\caption{\it 
Experiments studied here for probing the SD absorption signals.
}
\label{tab:SD:Exposure}
\end{center}
\end{table}
\begin{table}[th]
\renewcommand\arraystretch{1.58}
\begin{center}
\begin{tabular}{c c c c c c}
Isotope (Abund.) & $J$ & $\mathbb{S}_{p}$ & $\mathbb{S}_{n}$ & Ref.
\\\hline\hline
$ {}^{1}_{1}\mathrm{H} $ (99.985\%)  & 1/2 & 0.5 & 0 & \cite{Ellis:1987sh,Ellis:1991ef}
\\[2mm]\hline
$ {}^{6}_{3}\mathrm{Li} $ (7.5\%)  & 1/2 & 0.472 & 0.472 & \cite{Girlanda:2011fh,CRESST:2022dtl}
\\
$ {}^{7}_{3}\mathrm{Li} $ (92.5\%) & 3/2 & 0.497 & 0.004 & \cite{Pacheco:1989jz}
\\[2mm]\hline
$ {}^{13}_{\,\,\,6}\mathrm{C} $ (1.1\%)  & 1/2 & -0.009 & -0.172 & \cite{Engel:1989ix}
\\[2mm]\hline
$ {}^{19}_{\,\,\,9}\mathrm{F} $ (100\%)  & 1/2 & 0.475 & -0.0087 & \cite{Divari:2000dc}
\\[2mm]\hline
$ {}^{27}_{13}\mathrm{Al} $ (100\%)   & 5/2 & 0.343 & 0.0296 & \cite{Engel:1995gw}
\\[2mm]\hline
\begin{tabular}{c} 
$ {}^{129}_{\;\;54}\mathrm{Xe} $ (26.4\%) \\ 
$ {}^{131}_{\;\;54}\mathrm{Xe} $ (21.2\%) 
\end{tabular}
& 
\begin{tabular}{r} 1/2 \\  3/2  \end{tabular}
& 
\begin{tabular}{r} 0.0128 \\  -0.012  \end{tabular}
& 
\begin{tabular}{r} 0.300 \\   -0.217 \end{tabular}
&
\cite{Ressell:1997kx}
\\[2mm]
\hline\hline
\end{tabular}
\caption{\it 
Spin expectation values ($\mathbb{S}_N$) of the isotopes studied in this paper.
}
\label{tab:exps}
\end{center}
\end{table}
The readers can also find some averaged values in the Ref. \cite{DelNobile:2021wmp}.

\acknowledgments
K.M. was supported by the Natural Science Basic Research Program of Shaanxi (Program No. 2023-JC-YB-041).


\providecommand{\href}[2]{#2}\begingroup\raggedright\endgroup

\end{document}